\pdfoutput=1
\documentclass[aps,prl,twocolumn]{revtex4-1}
\usepackage{graphics}

\usepackage{color}
\usepackage{marvosym }
\graphicspath{{figs/}}

\usepackage{dsfont}
\usepackage{physics}
\usepackage{tabstackengine}
\stackMath
\usepackage{xcolor}

\usepackage{booktabs}
\usepackage{amsmath}
\usepackage{hyperref}
\usepackage{amssymb}
\usepackage{amsfonts}
\usepackage{wrapfig}
\usepackage{pstricks}
\usepackage{multirow}
\usepackage{bm}
\usepackage{dcolumn}
\newcommand{\etal}{\textit{et al.\ }}
\usepackage{mathtools}

\usepackage{booktabs}
\begin{document}
\title{Spin-polarized two-dimensional electron/hole gases on LiCoO$_2$ layers.}
\author{Santosh Kumar Radha and Walter R. L. Lambrecht}
\affiliation{Department of Physics, Case Western Reserve University, 10900 Euclid Avenue, Cleveland, OH-44106-7079}
\begin{abstract}
  First-principles calculations
  show the formation of a 2D spin polarized electron (hole) gas on the
  Li (CoO$_2$) terminated surfaces of finite slabs down to a  monolayer of
  LiCoO$_2$ in remarkable contrast with the bulk band structure stabilized
  by Li donating its
  electron to the CoO$_2$ layer forming a Co-$d-t_{2g}^6$ insulator.
  By mapping the first-principles computational  results to a minimal tight-binding models corresponding to a non-chiral 
  3D generalization of the quadripartite Su-Schriefer-Heeger (SSH4) model,
  we show that these surface states have  topological origin. 
\end{abstract}
\maketitle
LiCoO$_2$ has been mostly studied as cathode material in Li-ion batteries.\cite{Mizushima80,Miyoshi18,Iwaya13}
However, its layered structure also lends itself to the possibility of
extracting interesting ultrathin mono- or few layers nanoflakes.
A chemical exfoliation procedure has recently been established by
Pachuta \etal\cite{Pachuta} and similar exfoliation studies have also been done on Na$_x$CoO$_2$.\cite{Masuda06} The $R\bar{3}m$
structure of LiCoO$_2$ consists of
alternating CoO$_2$ layers,  which consist of edge sharing CoO$_6$ octahedra,
and Li layers stacked in an ABC stacking. By replacing lithium
by large organic ions, the distance between the layers swells and they can
then be exfoliated in solution and redeposited on a  substrate of choice
by precipitation with different salts. 


Inspired by these exfoliation experiments
we investigated the electronic structure of LiCoO$_2$ few layer
systems with various Li and other ion terminations and as function  of
thickness of the layers using density functional theory (DFT)
calculations.\cite{questaal-paper,compdetails}
In the process, we found, surprisingly that Li no longer fully donates its
electron to the CoO$_2$ layer but instead a surface state appears above
the Li and is occupied with a fraction of an electron ($\sim0.25$)
per Li. The amount of charge residing in this surface state was found
to be remarkably robust as function of thickness of the number of
LiCoO$_2$ layers, indicating
that this is a surface rather than ultrathin film effect.

As we will show, the Li bands in bulk LiCoO$_2$ lie at energies $E>5$ eV above
the Fermi level, consistent with the mostly ionic charge donation picture
mentioned above. So, the fact that a Li related surface state comes
down sufficiently close to the Fermi level to become partially occupied
is truly surprising. Furthermore because it is accompanied by the opposite
surface CoO$_2$ becoming spin-polarized it leads actually to a spin-polarized
electron gas on the Li side which is located primarily above the Li atoms.
Apart from the possibilities this may offer for interesting
physics, the main question we address in this paper is: why does this happen?
The answer we propose is that this is a topological effect. We show that
the DFT calculations can be explained by a minimal tight-binding (TB) model,
closely related to 
the quadripartite Su-Schrieffer-Heeger (SSH4) model which has been shown
to support topologically protected surface states for specific conditions
on the interatomic hopping integrals.\cite{Atherton16}
However, while the original SSH4 model has chiral symmetry protecting
the surface state at zero energy,  in the present case, the Li/CoO$_2$ electronegativity difference leads to surface states which would tend to
still place the surface electrons on the CoO$_2$ side. The crucial element
that allows the Li surface state to become partially filled is the strong lateral
interaction between Li atoms on the surface. The resulting band broadening
leads the Li surface band to dip below the top of the CoO$_2$ localized 
surface band leading to a partial electron/hole occupation
in these bands respectively.
As a further proof of the importance of the lateral interaction, we find
that when we place 1/2 Li per cell on opposite sides of the slab,
whereby the Li occur along 1D rows, the Li surface band has then only
1/3 of the band width and no longer dips below the Fermi level.

\begin{figure*}[!htb]
	\includegraphics[width=17cm]{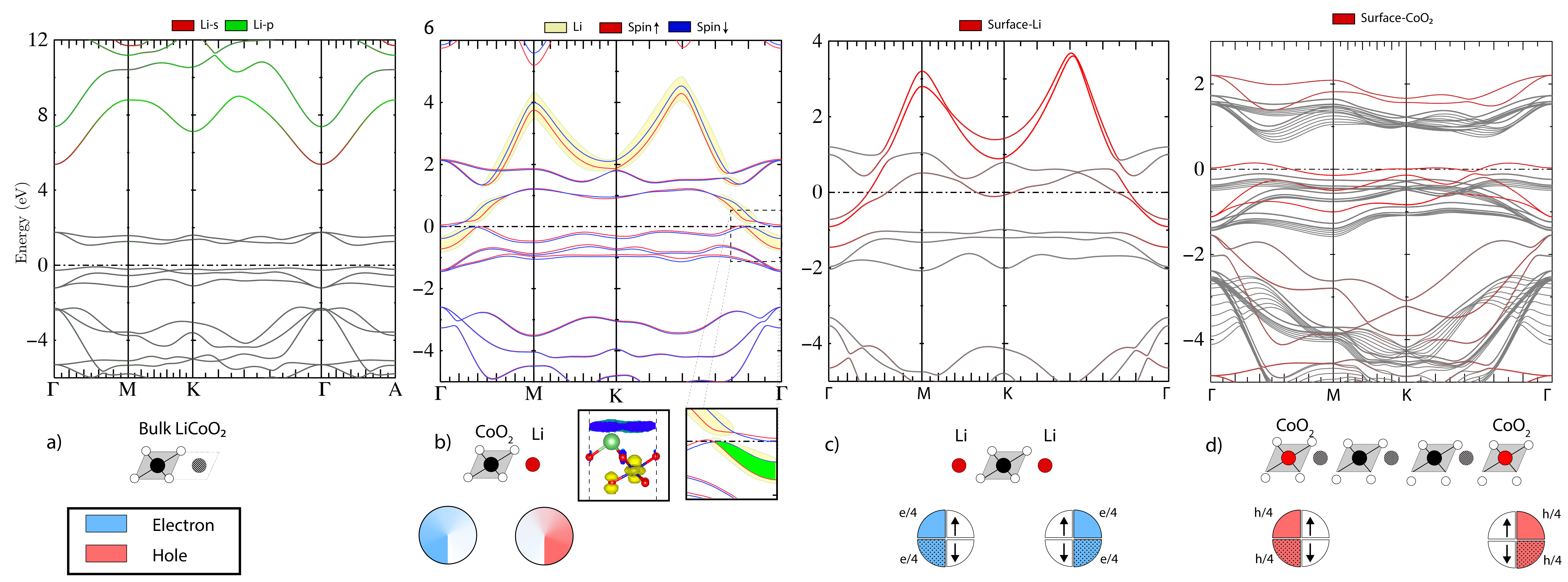}
	\caption{\textit{(a)} Bulk band structure of LiCoO$_2$,   \textit{(b)}
          Spin polarized  2D LiCoO$_2$ mono-layer. The \textit{green} region in the rightmost inset below (corresponding to the dashed line box in the band figure) shows the spin polarization in the 2DEG. The inset
          on its left shows the density of the wave function at $\Gamma$ of the Li band in dark blue.  \textit{(c)} Li terminated LiCoO$_2$ with 2D electron gas \textit{(d)} CoO$_2$ terminated LiCoO$_2$ with 2D hole gas. The circles colored red and blue blow the figures indicate the fractionalization of the charge in the surface states. In panels (c,d) it is determined by symmetry whereas in
          (b) it depends on the band parameters. The structural
          model corresponding to each case is also indicated below the corresponding band figure. \label{fig:mono_bands}}
\end{figure*}

We start by comparing the band structure in bulk LiCoO$_2$ with that of a monolayer LiCoO$_2$ in Fig. \ref{fig:mono_bands} (a) and (b). In the bulk case, we find an insulating band structure with a gap between the filled $t_{2g}$ and $e_g$ bands Co-$d$. The Li $s$ and $p$ derived bands, highlighted in color
occur at high energy indicating that they donate their electron to the
Co-$t_{2g}$ orbitals and support a mostly ionic picture of the bonding.
In strong contrast, in the monolayer system, we find an additional set of
spin-polarized bands, as highlighted
in the figure by  yellow shading.
By orbital decomposition, it is clear that this band
is Li related and its 2D dispersion closely matches that of a hypothetical 2D
monolayer of Li atoms. Importantly, we also find that it has not only Li-$s$
but also Li-$p_z$  character indicating the formation of Li-$sp_z$ hybrid
states. This free-electron-like band has clearly
avoided crossings with the Co-$e_g$ bands around 2 eV. It dips down
below the Fermi level with an electron pocket near $\Gamma$.
Inspection of  the corresponding wavefunction modulo squared shown in the inset
below it shows clearly that it is a surface state hovering slightly
above the Li atom. This band depends somewhat on the location chosen for
the Li atom, as shown in Supplemental Material (SM),\cite{SM}
but its general characteristics  are robust. In the lowest energy structure it
is found to be spin polarized as seen by the splitting of the up and down
spin bands and indicated by green shading in the inset below.
The fact that its maximum density lies above the Li atom
is consistent with a $sp_z$ hybrid orbital derived band. A comparison of the
bands for different Li locations in the surface unit cell is given
in SM.\cite{SM}
Further inspection of a planar average of the
the total electron density of the LiCoO$_2$ 
and integrating this in the $z$ region from just above the Li site into
the vacuum indicates a total charge of $\sim$0.25 $e$. 
This corresponds to a rather high  density  of  $5\times10^{14}$ $e$/cm$^{2}$. 
However we need to point out that this precise value depends on the
cut-off of the spatial region over which we integrated. 
Further calculations for thicker slabs with 3 and 4 layers with the same
termination of one Li terminated and one CoO$_2$ terminated surface
indicate that the surface charge density is remarkably robust.
Orbital decomposition in these thicker slabs
shows that a second surface state occurs
near the Fermi level  and is localized on the opposite CoO$_2$ terminated
surface layer and crosses the Li related one.
We find that there is a net hole concentration on that layer.

To rule out that this would be a supercell  artifact arising from the polar
nature of the structure, in which an artificial dipole could arise over
the vacuum region, we also consider a symmetrical slab with both surfaces
Li terminated. In this case, the system is in some sense overcompensated
by having one additional Li. The band structure for this case is shown
in Fig. \ref{fig:mono_bands}(c). 
A similar surface state is then found on both both Li terminated surfaces
and, in fact, one can see that the occupied electron pocket in this band
near $\Gamma$  is larger. We will later show that for symmetry  reasons
it must contain a fractionalized $e/2$ for each spin, so a net
charge of 1 electron per Li but for each spin it is in a single state
spread equally over both surfaces. Similarly, in the case of
a symmetric CoO$_2$ termination a partially filled 
surface state occurs on both surface layers with equal  hole concentration
by symmetry. 
This is shown in Fig. \ref{fig:mono_bands}(d). 

We also considered a 1/2 Li per Co symmetrically at both terminating
surfaces.  In that case the Li is placed in 1D rows on the surface and
a 1D electron gas is found above these Li rows, as shown in SM.\cite{SM}
However, the density of electrons in this case is smaller by about a factor 10. 
We further inspect the dilute limit of 1 Li per 4 Co atoms
and still find an even smaller residual small charge density in an
orbital locally above that Li. However, no Li localized
surface state dipping below the Fermi level is found in these cases with
a reduced Li surface concentration.  This indicates that sufficient
lateral interaction between Li atoms is required to generate a
significant occupation of the surface states with electrons.
Replacing the Li terminating layer by Be (also overcompensating the system
from the CoO$_2$ point of view) we find a higher electron density in a Be
related surface band. Replacing Li by Na gives similar results but
with different band widths of the surface band because of the
stronger overlap of the Na orbitals.  The band structures of
these cases are all shown  in SM.\cite{SM}.

To explain these remarkable results, we now consider a minimal tight-binding
model.  First, it is clear that the Li needs to be represented by two $sp_z$
orbitals pointing toward the CoO$_2$ layer on either side. It is well known
that an even number of orbitals is required in a 1D model to obtain
topologically non-trivial band structures.
Therefore we choose to represent
the CoO$_2$ layer by two $s$-like Wannier orbitals. One could think of these
as representing the $a_1$-symmetry of the $D_{3d}$ group or $d_{z^2}$ orbitals
on Co with $z$ along the layer stacking ${\bf c}$-axis making bonding
orbitals with O-$p_z$ on either side of the Co.  Of course,
this does not represent the full set of CoO$_2$ layer derived bands but
we will argue that it represents the relevant bands leading to the
surface states. The important point is that the CoO$_2$ and Li
each are represented by two Wannier type orbitals whose centers are
not on the atoms but on the bonds in between atoms in the layer stacking
direction.

This minimal model is then a non-chiral version of the
SSH4 model. Ordering the orbitals as  $\{\ket{Li^a},\ket{CoO_2^a},\ket{Li^b},\ket{CoO_2^b}\}$ the Hamiltonian for the above 1D system (with distance between the layers set to 1) is represented by the following $4\times4$ matrix:
\begin{eqnarray}
H_{1d}&=&\begin{pmatrix}
\delta & 0 & \tau_1&\tau_4 e^{ik_z} \\ 
0 & -\delta& \tau_2&\tau_3 \\ 
\tau_1 &\tau_2  & \delta& 0\\ 
\tau_4e^{-ik_z} &\tau_3  &0  &-\delta 
\end{pmatrix}, \label{eq:1}\\
&=&\begin{pmatrix}
\delta\sigma_z & {\bf s}^*(k_z) \\
      {\bf s}(k_z)&\delta\sigma_z
\end{pmatrix} \label{eq:2} \\
&=&\sigma_x\otimes{\bf h}(k_z)-i\sigma_y\otimes {\bf a}(k_z)+\delta \mathds{1}_2\otimes \sigma_z\label{eq:3}
\end{eqnarray}
where $\tau_1$,$\tau_2$,$\tau_3$, are intra-unit cell interaction while
$\tau_4=\tau_2$ is the out of unit cell interaction, $\delta$ is the ionic on-site term for Li relative to CoO$_2$. The second form of the
Hamiltonian focuses on its $2\times2$ block structure, in which
${\bf s}(k_z)$ is a $2\cross2$ matrix
which is split in its hermitian, ${\bf h}(k_z)={\bf h}(k_z)^\dag$, and
anti-hermitian, ${\bf a}(k_z)^\dag=-{\bf a}(k_z)$, parts, allowing us to finally
write the block structure of the Hamiltonian in terms of the Pauli matrices
and a $2\times2$ unit matrix $\mathds{1}_2=\sigma_0$. 
For our system, $\tau_2=\tau_4=t^z_\mathrm{Li-CoO_2}$ corresponds to the interaction between the  Li and CoO$_2$ layers while $\tau_1$=$t^z_{Li}=(E_s^{\mathrm{Li}}-E_p^{\mathrm{Li}})/2$ corresponds to the interaction between the two Li $sp_z$'s on the same Li atom, and $\tau_3=t^z_\mathrm{CoO_2}$  to O-Co-O interaction within the layer. 

This model, which is the SSH4 model for $\delta=0$ corresponding to chiral
symmetry, has  been shown \cite{Atherton16} to have non-trivial topology which
requires zero-energy  edge states when $\tau_1\tau_3<\tau_2\tau_4$.
In fact, in that case, the winding number,
which characterizes the topology $\mathcal{W}=\oint \frac{dk_z}{2\pi}\partial_{k_z}\arg{\det{s(k_z)}}$ is 1, while in the other case it is 0.
This condition in our case, indicates that the covalent Li-$sp_z$--CoO$_2$
interaction is stronger than the intra CoO$_2$ interaction or the
Li-$sp_z$ interaction on the same Li atom.

When $\delta$ is not zero this model becomes non-chiral and the zero energy
surface states move up and down in energy and become
localized on opposite edges which would tend to
localize the electrons  on one side only, in the present case obviously
the CoO$_2$ side because it would have lower energy $-\delta$ because of
its electronegative character.  Therefore, to explain the electron
occupation of the Li-derived surface state, we need to generalize our
model to include the lateral in-plane interactions.

\begin{figure*}[!htb]
	\includegraphics[width=17.5cm]{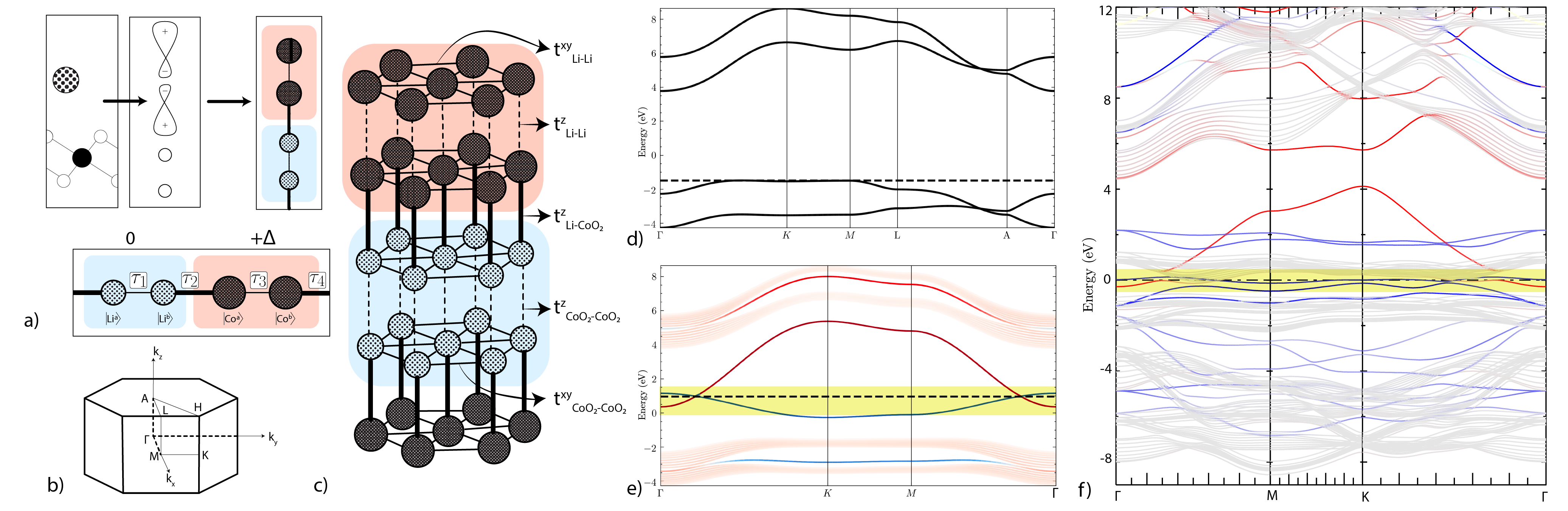} 
	\caption{a) LiCoO$_2$ unit cell along with orbitals after hybridization and the final model having all relevant physics b) 3D Hamiltonian considered for calculation and the BZ c) 3D model of the Hamiltonian and the corresponding hopping terms d) Bulk band of the 3D Hamiltonian. e) Band structure of a finite slab extending in $x-y$ direction. \textit{red} color corresponds to the edge Li atom while \textit{blue} to bottom most CoO$_2$ f)  Spin-less DFT band structure of 14nm unit cell with Li terminated on one side and CoO$_2$ layer on another with colors representing the same as in previous figure. \label{fig:TBbands}}. 
\end{figure*}

\begin{figure}
 	\includegraphics[width=7cm]{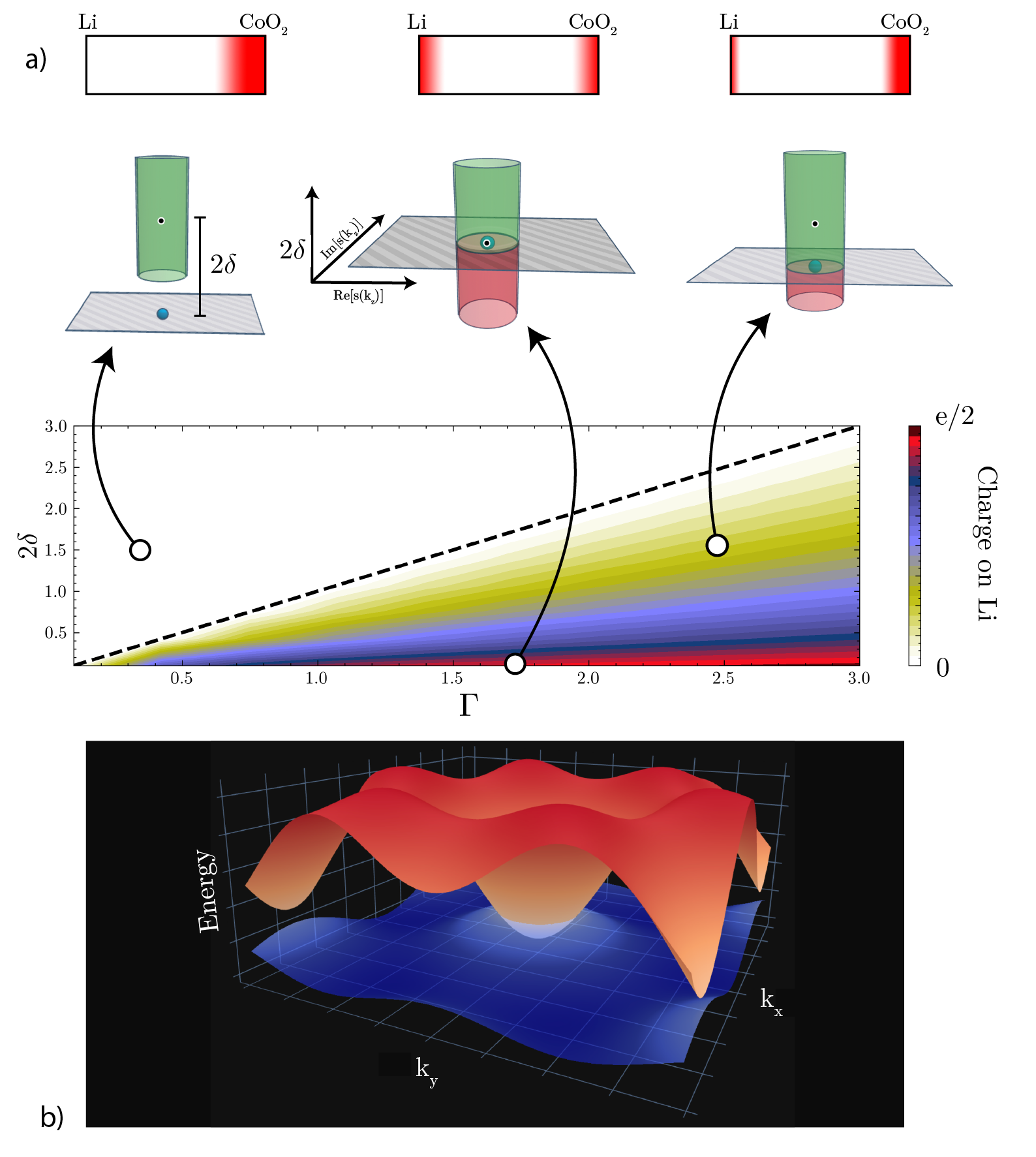} 
 	\caption{(a) Schematic representation of Li and CoO$_2$ surface charge \ for different choices of
          model parameters, shown below in 
          (b) cylinder topological model, (c) surface charge density on Li side (as color map) per Li calculated within the full TB model 
          as  function of on-site term $\delta$ and in-plane
          band width $\Gamma$, (d) 3D view of intersecting surface bands.\label{fig:u-vs-v}}
\end{figure}

We introduce in-plane $t^{xy}_\mathrm{Li}$ and $t^{xy}_\mathrm{CoO_2}$
interactions on the planar trigonal lattice 
\begin{equation}
f_{Li(Co)}=2t^{xy}_{Li(Co)}\sum_{i=1}^{3} cos(\boldsymbol{k}\cdot\boldsymbol{\delta_i}) 
\end{equation}
where $\pm \boldsymbol{\delta}_i$ are the $6$ vectors pointing toward  
the nearest neighbors and $\boldsymbol{k}_\parallel=k_1{\bf b}_1+k_2{\bf b}_2=(k_x,k_y)$ is the in-pane 2D wave vector. 
This modifies only the diagonal terms in \eqref{eq:2}  leading to 
\begin{eqnarray}
  H_{3d}(k_x,k_y,k_z)&=&\sigma_x\otimes {\bf h}(k_z)-i\sigma_y\otimes{\bf a}(k_z) \nonumber \\
 &&  +\sigma_0\otimes[\delta+\Delta({\bf k}_\parallel)]\sigma_z\label{eq:6}
\end{eqnarray}
after we drop out a constant from the Hamiltonian diagonal. Here, 
$\Delta({\bf k}_\parallel)=\frac{f_{Li}-f_{Co}}{2}$,
is added to the $\delta$  in the 1D model and can be thought of as a dimensional crossover parameter,\cite{obana} which tunes the influence of the
in-plane dimensions.  Physically, $\Delta({\bf k}_\parallel)$
is proportional to the width of the energy bands in ${\bf k}_\parallel$
space, which is $\Gamma=\Gamma_{Li}+\Gamma_{CoO_2}=6(t^{xy}_\mathrm{Li}+t^{xy}_\mathrm{CoO_2})$.

\autoref{fig:TBbands} (c) shows the 3D- layered tight-binding
model and the corresponding hopping
parameters described above.   \autoref{fig:TBbands} (d) shows the band structure of the above $3D$ Hamiltonian  while (e) shows the energy levels of the
$2D$ periodic system with a finite number of layers along the $z-$axis with top most layer having Li  and bottom layer CoO$_2$. The bands are color-weighted (Li-red, Co-blue).
While the 3D periodic TB system is seen to have a wide
gap between high-lying Li derived bands and low lying CoO$_2$-derived
bands, two surface bands with significant ${\bf k}_\parallel$ dispersion
are seen in part(e), which are respectively localized on
the Li (red) and CoO$_2$ sides (blue) and are found to cross each other
near the Fermi energy.
In fact, we can see that besides the surface bands
lying in the overall 3D gap, two more states have significant weight
on the end Li and CoO$_2$ orbitals but they lie within the range of
the bulk bands occurring in a gap in the projected bands but not in the
overall gap between occupied and empty states.

The interlayer parameters
used in the TB Hamiltonian are chosen to satisfy the SSH4
non-triviality criterion. The in plane interactions determining
$\Delta({\bf k}_\parallel)$
are chosen to resemble the DFT band structure (shown in part (f) for a 14 nm
thick (in the $z$-direction) in-plane periodic layer)
and turn out to satisfy $t^{xy}_\mathrm{CoO_2}\approx -0.1t^{xy}_\mathrm{Li}$. The opposite
dispersion of these surface bands is obvious from the DFT results
and translates to these in-plane directions having opposite sign.
In the actual system, it is clear that
$t^{xy}_\mathrm{Li} <0$ as it is a $\pi$-interaction
between Li-$p_z$ states combined with $\sigma$-interaction between the $s$-part of
the $sp_z$ orbitals and in absolute value
is much larger than the  $t^{xy}_\mathrm{CoO_2}>0$.
This is important because it means that the Fermi level is pinned
at the intersection of the two surface bands, which indicate an overall
semimetallic case with as many holes in the CoO$_2$ surface band
as there are electrons in the Li surface band when they overlap.

While we recognize that this model is representing only part of the
bands of the actual physical system, the correspondence of the
surface bands in the DFT and the TB-model Hamiltonian
is convincing that it captures the essence of the
relevant physics. Essentially, we thus conclude that
the surface states originate from
the topologically nontrivial SSH4 character of the interlayer bonds
which correspond to a non-zero winding number. To this is added a
term in the Hamiltonian orthogonal to the space of the $\sigma_x$,
$\sigma_y$ parts of the Hamiltonian which define a complex plane in
which the winding number is defined. One can generally write such
a Hamiltonian as $H={\bf d}\cdot\bm{\sigma}$ where $d_\parallel$ corresponds
to the $(x,y)$ components in a complex plane defining the winding
number.\cite{Pershoguba12,Mong}
The loop defining the winding number is now above or below the
complex plane. Its projection on the complex plane encircling the origin
or not defines the winding number and therefore topological non-trivial/trivial
character. The component $d_\perp$ to this plane determines the energy
position of the surface states $E_s=\pm|d_\perp|$,\cite{Mong}
away from zero by the chiral symmetry breaking.
In the bipartite SSH model, the third component
of the $\bm{\sigma}$ would be simply the third Pauli
matrix $\sigma_z$ while in our case it is $\sigma_0\otimes\sigma_z$.
However, because of the in-plane band dispersion, such an  SSH4 like model
now applies at each ${\bf k}_\parallel$. This turns the loop outside the plane
into a cylinder (shown in Fig. \ref{fig:u-vs-v}(b)
centered at energy $2\delta$ from the plane
with height given by the 2D band width.

The amount of charge 
on the Li is determined (shown in Fig. \autoref{fig:u-vs-v}(a,c) as function
of the model parameters) by how much the bottom of the Li band overlaps
with the top of the CoO$_2$ band, as shown in Fig. \autoref{fig:u-vs-v}(d).
Assuming a steplike density of states (DOS) for each band near
these band edges and parabolic free electron like bands, which
makes sense near the band edges for a 2D system, we can easily determine
the Fermi energy, located in the surface bands from the fact that the
number of electrons in the upper band equals the number of holes
in the lower band. Using the density of states for free electrons in 2D
to be proportional to the effective mass in each band, 
and the proportionality of the inverse effective mass to the hopping
parameter or bandwidth of the  tight-binding model for that band, we
find that 
\begin{equation}
\frac{\text{charge on Li-side}}{\text{charge on CoO$_2$-side}}=
\begin{cases} \frac{\Gamma -2\delta}{\Gamma +2\delta},& 2\delta\le\Gamma \\
0&2\delta\ge\Gamma\label{eq:7}
\end{cases}
\end{equation}
One can easily see that if the splitting of the two surface band centers
(which is $2\delta$) is larger than the sum of half their band widths
then the charge will still all be
localized on the CoO$_2$ and zero on Li.
A full numerical calculation of the net surface charge density within
the tight-binding model resulting from the overlapping bands is
given in SM and shown to closely agree with the above approximate
result.\cite{SM}
In the cylinder topological model mentioned above, the part of the
cylinder that dips below the surface corresponding to $\delta=0$
gives the amount of charge on the Li side, again when assuming
a constant DOS. 

The above model is consistent with the facts from our DFT calculations
presented in the SM, that for Na with larger in-plane
interactions and hence larger 2D surface
band width, a larger surface charge density is found than for Li on the
surface. The same is true for Be which also has a smaller electronegativity
difference and hence smaller $\delta$ in our model and, in fact gives
an additional electron to the surface states.
Finally, when two equal surface terminations are used then there is
an overall inversion/mirror symmetry in the center of the slab and thus
by symmetry requires that the additional electron in the surface states
is spread equally over both sides. The same is true for holes for the
case of two CoO$_2$ terminated surfaces.

In the SM, we show furthermore that the occurrence of the surface states
is related to the entanglement of the states in the two halves of the system
that are separated by creating the surface by calculating
the entanglement spectrum.\cite{Fidkowski,Turner10,Calabrese_2016,Alexandradinata}


In summary, we have shown that surfaces of LiCoO$_2$ finite slabs
host  topologically required surface states related to the SSH4 like
nontrivial interlayer interactions of Li-$sp_z$ and CoO$_2$
bond-centered Wannier orbitals. As a  result of strong lateral
interactions, the Li related surface band can become partially occupied
and host a spin-polarized 2DEG of fairly high electron density. 
While several angular resolved
electron spectroscopy (ARPES) and scanning tunneling microscopy
studies (STM) have been published in the past \cite{Yang07,Yang05,Qian06,Shimojima06,Hong19,Iwaya13} for both
Li$_x$CoO$_2$ and Na$_x$CoO$_2$ they were generally focused
on the bulk rather than on the search for surface states, which may thus have been missed.

{\bf Acknowledgements:} This work was supported by the U.S. Air Force Office
  of Scientific Research under Grant No. FA9550-18-1-0030.  The calculations made use of the High Performance Computing Resource in the Core Facility for Advanced Research Computing at Case Western Reserve University.
\bibliography{lmto,dft,licoo2,top}

\section{Supplemental Material}
\subsection{Density functional theory  results.}
In this section of the supplemental material, we provide additional
information on various density functional theory results.
\begin{figure}[!htb]
	\includegraphics[width=7.5cm]{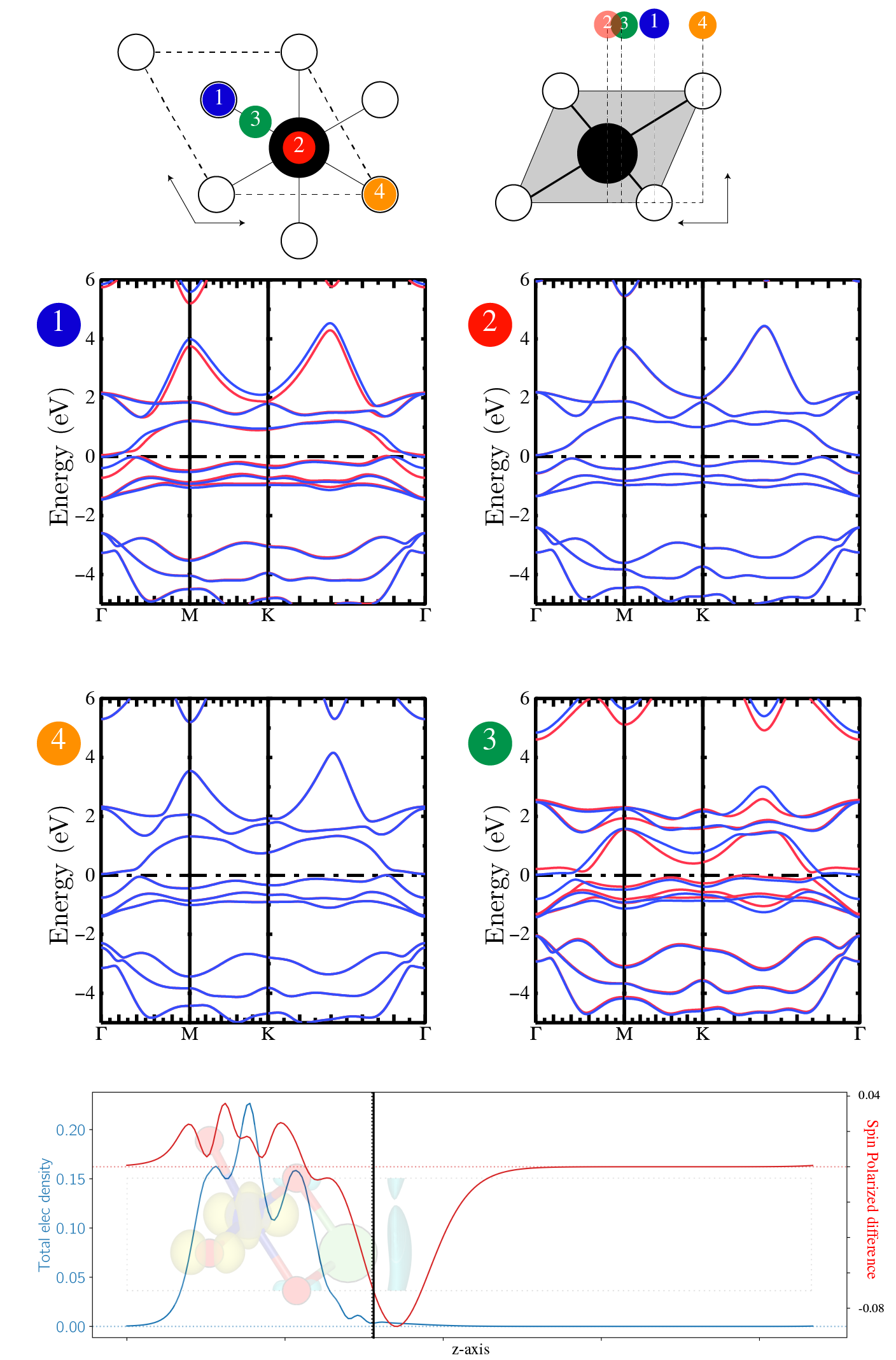} 
	\caption{\textit{(top)} top and side view of all 4 symmetric positions of Li on top of 2D CoO$_2$ lattice. \textit{(middle)} Spin polarized LDA band structures for the corresponding structures. \textit{(bottom)} Total electron density (blue) integrated along in-plane axis and spin difference =$n_{\downarrow}-n_{\uparrow}$ (red). Black line shows the real space limit for integrating the surface charge density  \label{fig:li_placement}}. 
\end{figure}

\begin{figure*}[!htb]
	\includegraphics[width=15cm]{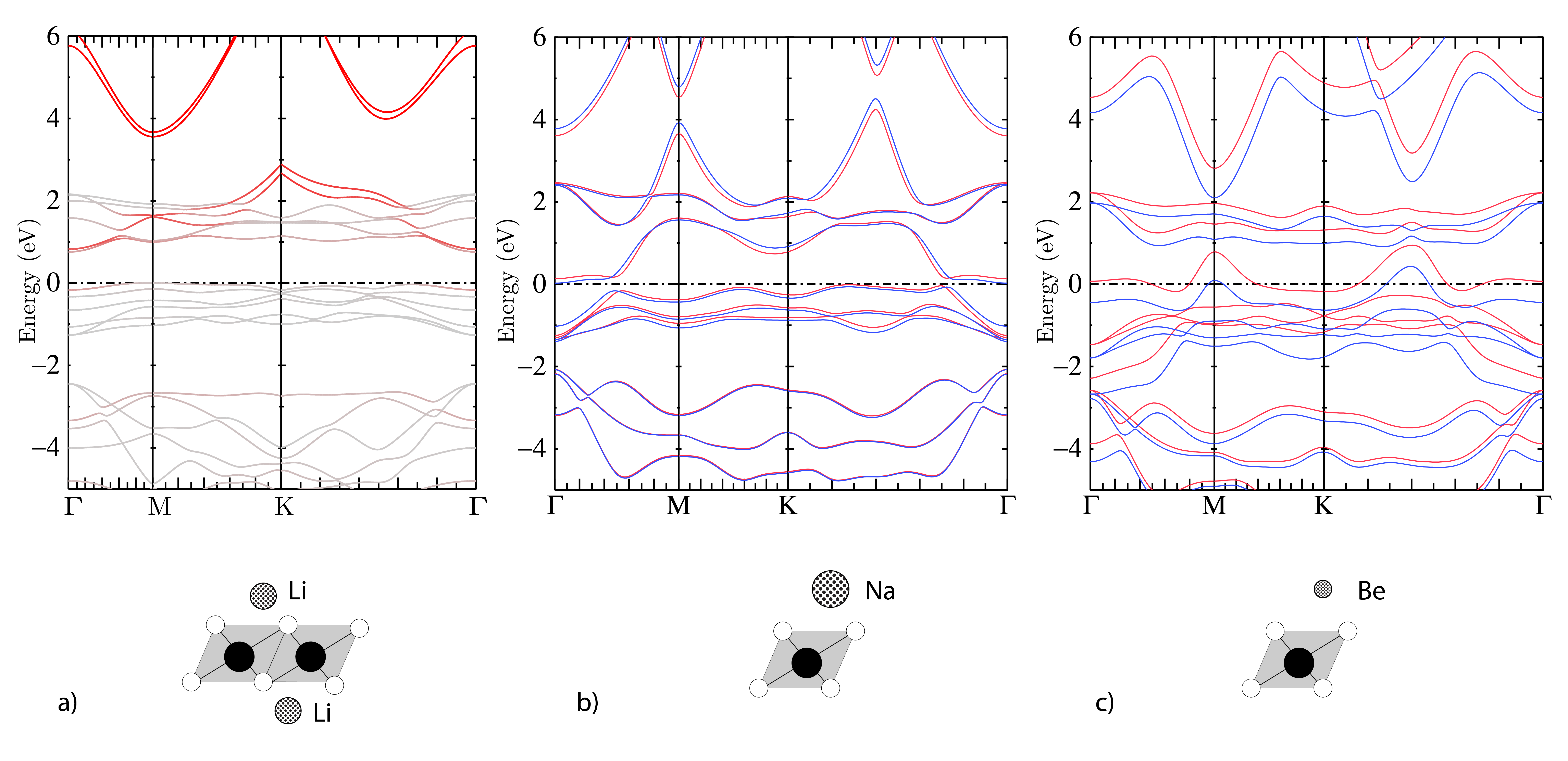} 
	\caption{ (a) LiCoO$_2$ mono-layer with the Li atoms forming 1D chain (Red color is the Li projected band); (b) NaCoO$_2$; (c)  BeCoO$_2$ band structures (\textit{red/blue} bands denote the majority and minority spins). \label{fig:other_atoms}}. 
\end{figure*}

In Fig. \ref{fig:li_placement} we show the changes in band structure for different surface locations of the Li.   Their relative total energies are given 
in Table \ref{tab:totalE} . 
The lowest energy position  \textbf{(1)} for Li is on top of the Co, position
\textbf{(2)} is
above the bottom O,  \textbf{(4)} above the top layer O and  \textbf{(3)} in the center of the
2D cell.

\begin{table}[!htb]
	\centering
	\caption{Relative energy of the structures in \autoref{fig:li_placement} } \begin{ruledtabular}
		\begin{tabular}{@{}lllll@{}}
			Structure   & 1 & 2    & 3    & 4    \\ \hline
			Energy (eV/f.u.) & 0 & 0.09 & 0.32 & 0.81 \\ 
		\end{tabular}
	\end{ruledtabular}
	\label{tab:totalE}
\end{table}

It is interesting to note that in structure \textbf{(3)} shown in \autoref{fig:li_placement}, the effective interaction is between Li and CoO$_2$ layer is mediated by O-$p$ instead of Co-$d$. This change is captured in the band structure by the lowering of the center of the \textit{free-electron} like Li band  compared to other cases. On the other hand, this is clearly
not the lowest total energy.
Interestingly, we find negligible spin-polarization
for both locations  \textbf{(2)} and  \textbf{(4)}.

\begin{figure}
	\includegraphics[width=8cm]{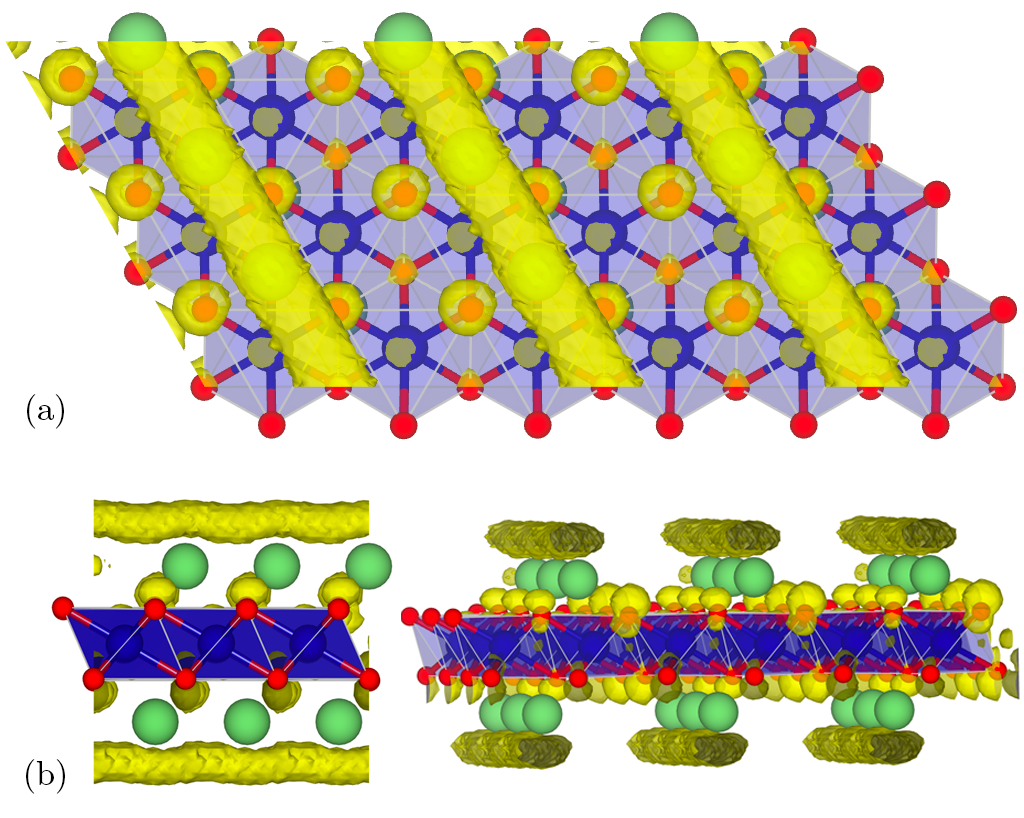}
	\caption{2D electron density in case of CoO$_2$ monolayer
		with Li on either side but arranged in rows of Li, achieving
		one Li per CoO$_2$. yellow isosurface
		of the electron density
		correspond to $1.04\cross10^{-5}$ .
		\label{fig:1deg}}
\end{figure}
\begin{figure}
	\includegraphics[width=8cm]{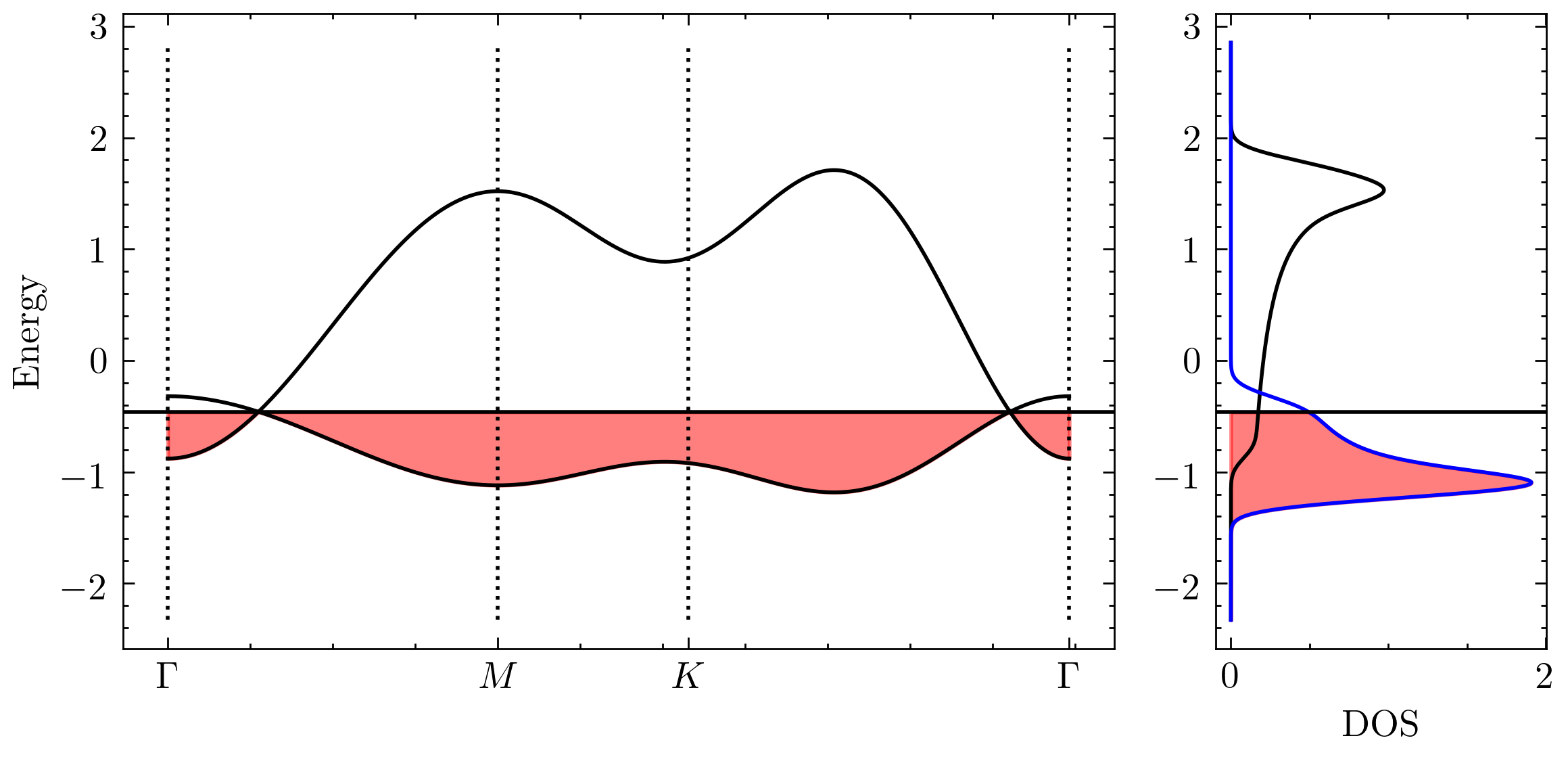}
	\includegraphics[width=8cm]{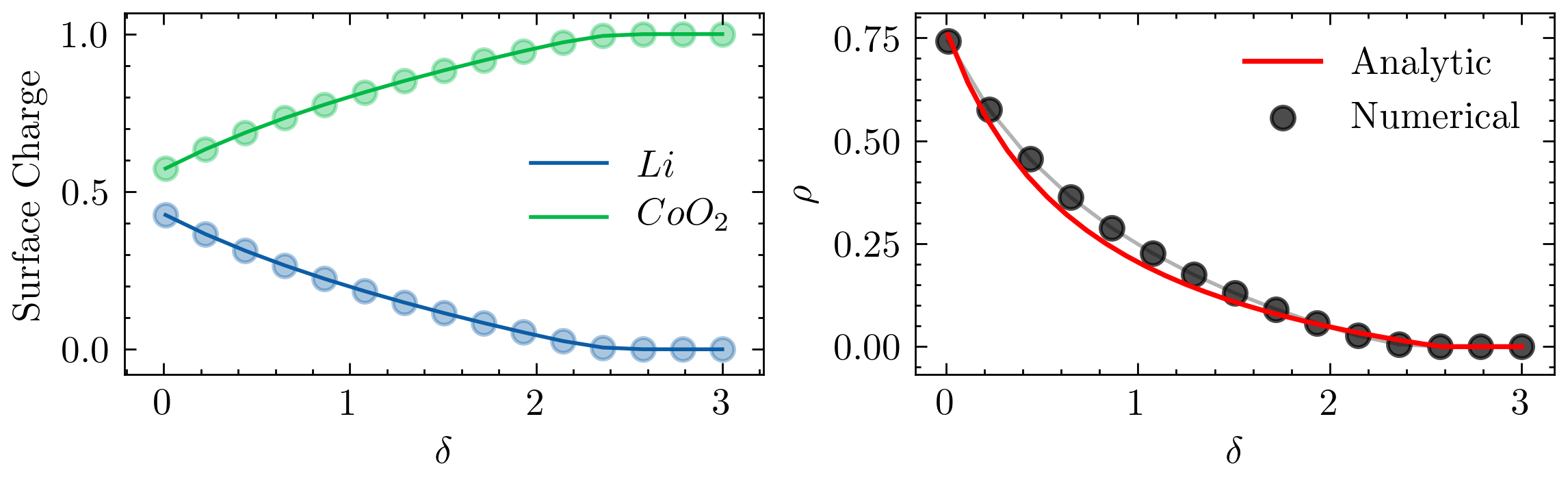}
	\caption{\textit{(Top)} Low energy band structure of surface states and Density of States (DOS) contribution from each band (black on Li side, and
		blue on CoO$_2$ side)
		Surface states have a 2D dispersion corresponding to
		a triangular lattice with nearest neighbor  hopping parameters 
		given by $t^{xy}_{CoO_2}=0.1$ and $t^{xy}_{Li}=-0.3$ in arbitrary
		units with an on-site offset of $\delta=0.9$  (in units of $\Gamma/6=(t^{xy}_{CoO_2}+t^{xy}_{Li})$), \textit{(Bottom left)} Surface charge
		per unit cell area as  function of on-site parameter $\delta$;  
		\textit{(Bottom right)} Ratio of charge on Li to charge on CoO$_2$
		surface layers from numerical TB and analytic equation Eq.(6) derived in the main text. In the limit of $\delta=0$, this ratio
		approaches 1.\label{fig:dos}}
\end{figure}

At the bottom of Fig. \ref{fig:li_placement} we show the 2D planar
averaged electron density and its spin polarization for location  \textbf{(1)}
set against the structure.
The region over which we integrated the surface density is to the right
of the vertical black line.

Next  we show the band structures for
various other cases in Fig. \ref{fig:other_atoms}. Part (a)  
shows the case of a monolayer of CoO$_2$ compensated by one Li per Co
but with Li arranged at half the surface density on each surface. The
Li atoms then occur in rows. The corresponding electron density is shown
in Fig. \ref{fig:1deg}. While showing some electron density
just above the Li atom rows, it should be pointed out that this
electron density is a factor 10 times smaller than for the full Li
coverage. Correspondingly, we see that the surface bands related to Li
do not dip below the Fermi level. Although centered at about the same energy,
the band width of this surface band is now about 3 times smaller because
the Li only have 2 neighbors (along the 1D rows) instead of 6 in the plane.
This prevents this surface band to become occupied. 
However, from the color coding  (red for
the Li contribution to the band) we can see that the highest occupied
band does contain some Li contribution and forms an electron pocket
around $\Gamma$.  The plot of the corresponding wave function
modulo squared is what is shown in Fig. \ref{fig:1deg}.  This Li
row related 1DEG cannot be explained within the SSH4 based tight-binding
model. It would require a more complete description of the
Li-CoO$_2$ layer interactions.

Next, in Fig. \ref{fig:other_atoms}(b) we show the case of a fully
Na covered CoO$_2$ monolayer with Co on one side. This is similar to
the corresponding Li case discussed in the main part of the paper but
show that with Na, the electron pocket around $\Gamma$ is increased in size.
This is consistent with the larger lateral interactions between Na.
Finally, in Fig. \ref{fig:other_atoms}(c) we show the case of Be covered CoO$_2$. Compared to Li, we now
overcompensate the CoO$_2$. In this case the electron density in the surface
2DEG is even larger but we also obtain a larger spin-splitting.

\subsection{Tight-binding model surface charge calculation} \label{app:chargecalc}
In Fig. \ref{fig:dos} we show the results of a numerical calculation
of the surface state occupancy as function of the energy separation
of the two surface bands within the tight-binding approximation.
We can see that it is in good agreement
with the model results in the main paper.
Note that beyond $\delta=3$ here the overlap of the bands
is zero and no charge occurs on the Li side. For $\delta=0$ we are
in the limit where the charge on the Li is 1/2 by symmetry. 

\subsection{Entanglement spectrum}
To understand the surface states better, we use the idea of \textit{entanglement spectrum} (ES)\cite{Li} which has been found to be a  generally useful
theoretical tool in investigations of topological states\cite{Fidkowski,Turner10,Calabrese_2016,Alexandradinata}.
The main idea of the ES is that the eigenvalues of the hermitian
correlation matrix of the occupied eigenstates, restricted to a subsystem
A of the combined system (A+B),  provide already information on the
existence of surface states when the system would be split in separate
A and B parts and of the topologically non-trivial nature of the system. 
In our case of non-interacting electrons,  the correlation matrix is defined in terms of the Bloch
functions expanded in the tight-binding basis set as follows. Although our system is periodic in $x$ and $y$ direction
we here consider Bloch states only in one direction combined with the layer direction $z$ in which the
non-trivial SSH4 topology applies. 
Let the eigenstates be  $|\psi^n_{k_x}\rangle=e^{i k_xx}|u_{nk_x}\rangle=\sum_{j\alpha} e^{ik_xx}[u^n_{k_x}]_{j\alpha} |\phi_{j\alpha}\rangle$,
where $i$ labels the sites, which can be either in the A or B  part of the system and $\alpha$ labels orbitals per site, 
then the correlation matrix restricted to the A-subsystems is given by
\begin{equation}
C^A_{i\alpha,j\beta}(k_x)=\sum_n^{occ} [u^n_{k_x}]_{i\alpha}^*[u^n_{k_x}]_{j\beta}, \qquad  \hbox{with\ } {i\in A}, {j\in A} 
\end{equation}
If we remove the restrictions on $i,j$ then we drop the superscript $A$.  The eigenvalues of this correlation matrix
$\xi(k_x)$ define the ES. If we would not include the restriction, this correlation matrix is built from idempotent
projection operators and thus has eigenvalues 0 or 1 only. As shown in \cite{Alexandradinata} and elsewhere, the existence
of eigenvalues deviating strongly from 0 or 1, near 1/2 are an indicator of the entanglement of the states
between its subparts and thus of the non-trivial topology and the existence of surface states. 
\begin{figure}[!htb]
	\includegraphics[width=8cm]{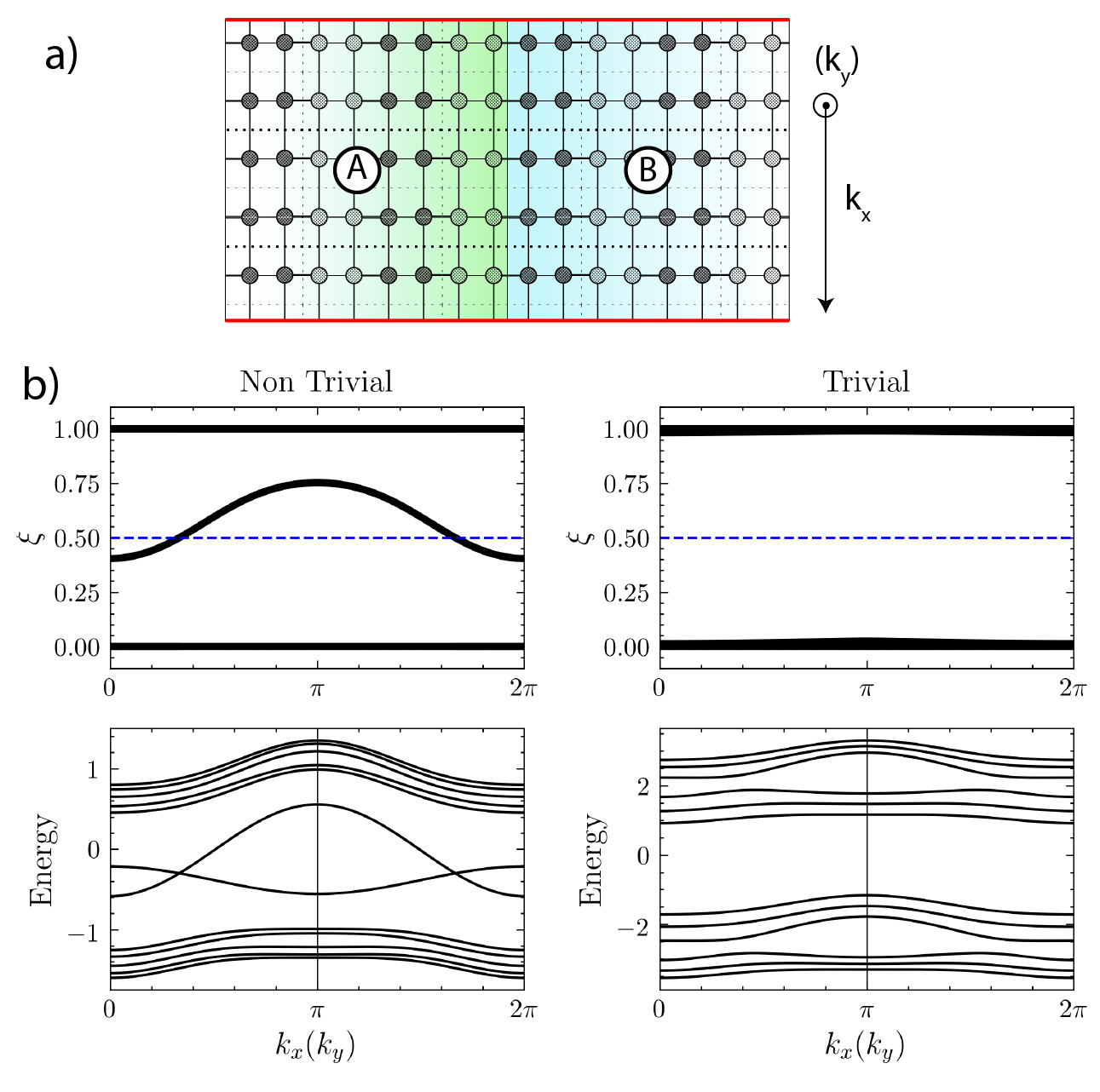} 
	\caption{a) Partitioning of the system in two parts for calculating the entanglement spectrum of the reduced 2D system   b) \textit{(top)} entanglement spectrum of the 2D system along the cut \textit{(bottom)} band structure of 1D ribbon of the corresponding 2D Hamiltonian showing the edge states
		for (left) topologically non-trivial and right trivial choice of SSH4 parameters.\label{fig:QE}}
\end{figure}

The eigenvalues $\xi(k_x)$ of  $C_{ij}(k_x)$ matrix at a given $k_x$ are shown in
\autoref{fig:QE}(b) along with the 1D eigenvalues of our tight-binding model  and clearly show the one-to-one correspondence
between the ES containing eigenvalues near 1/2 with the existence of surface states. 
Furthermore we see that for a different choice of the SSH4 TB parameters,
no surface states exist and correspondingly no ES with eigenvalues
near 1/2. 

The connection between $\xi$ and the previous topological picture
in terms of cylinders in Fig. 3 of the main paper, 
is that for each ${\bf k}_\parallel$ the eigenvalues  $\xi>0.5 \text{ }(\xi<0.5)$
correspond to a loop above (below) the $z=0$ plane in Fig. 3 of the main
paper as indicated by the \textit{green/red} color in the cylinders
of Fig. 3(a). The $\xi=0.5$ eigenvalue corresponds exactly to
the $z=0$ plane. 

\bibliography{lmto,dft,licoo2,top}

\begin{thebibliography}{21}%
\makeatletter
\providecommand \@ifxundefined [1]{%
 \@ifx{#1\undefined}
}%
\providecommand \@ifnum [1]{%
 \ifnum #1\expandafter \@firstoftwo
 \else \expandafter \@secondoftwo
 \fi
}%
\providecommand \@ifx [1]{%
 \ifx #1\expandafter \@firstoftwo
 \else \expandafter \@secondoftwo
 \fi
}%
\providecommand \natexlab [1]{#1}%
\providecommand \enquote  [1]{``#1''}%
\providecommand \bibnamefont  [1]{#1}%
\providecommand \bibfnamefont [1]{#1}%
\providecommand \citenamefont [1]{#1}%
\providecommand \href@noop [0]{\@secondoftwo}%
\providecommand \href [0]{\begingroup \@sanitize@url \@href}%
\providecommand \@href[1]{\@@startlink{#1}\@@href}%
\providecommand \@@href[1]{\endgroup#1\@@endlink}%
\providecommand \@sanitize@url [0]{\catcode `\\12\catcode `\$12\catcode
  `\&12\catcode `\#12\catcode `\^12\catcode `\_12\catcode `\%12\relax}%
\providecommand \@@startlink[1]{}%
\providecommand \@@endlink[0]{}%
\providecommand \url  [0]{\begingroup\@sanitize@url \@url }%
\providecommand \@url [1]{\endgroup\@href {#1}{\urlprefix }}%
\providecommand \urlprefix  [0]{URL }%
\providecommand \Eprint [0]{\href }%
\providecommand \doibase [0]{http://dx.doi.org/}%
\providecommand \selectlanguage [0]{\@gobble}%
\providecommand \bibinfo  [0]{\@secondoftwo}%
\providecommand \bibfield  [0]{\@secondoftwo}%
\providecommand \translation [1]{[#1]}%
\providecommand \BibitemOpen [0]{}%
\providecommand \bibitemStop [0]{}%
\providecommand \bibitemNoStop [0]{.\EOS\space}%
\providecommand \EOS [0]{\spacefactor3000\relax}%
\providecommand \BibitemShut  [1]{\csname bibitem#1\endcsname}%
\let\auto@bib@innerbib\@empty
\bibitem [{\citenamefont {Mizushima}\ \emph {et~al.}(1980)\citenamefont
  {Mizushima}, \citenamefont {Jones}, \citenamefont {Wiseman},\ and\
  \citenamefont {Goodenough}}]{Mizushima80}%
  \BibitemOpen
  \bibfield  {author} {\bibinfo {author} {\bibfnamefont {K.}~\bibnamefont
  {Mizushima}}, \bibinfo {author} {\bibfnamefont {P.}~\bibnamefont {Jones}},
  \bibinfo {author} {\bibfnamefont {P.}~\bibnamefont {Wiseman}}, \ and\
  \bibinfo {author} {\bibfnamefont {J.}~\bibnamefont {Goodenough}},\ }\href
  {\doibase https://doi.org/10.1016/0025-5408(80)90012-4} {\bibfield  {journal}
  {\bibinfo  {journal} {Materials Research Bulletin}\ }\textbf {\bibinfo
  {volume} {15}},\ \bibinfo {pages} {783 } (\bibinfo {year}
  {1980})}\BibitemShut {NoStop}%
\bibitem [{\citenamefont {Miyoshi}\ \emph {et~al.}(2018)\citenamefont
  {Miyoshi}, \citenamefont {Manami}, \citenamefont {Sasai}, \citenamefont
  {Nishigori},\ and\ \citenamefont {Takeuchi}}]{Miyoshi18}%
  \BibitemOpen
  \bibfield  {author} {\bibinfo {author} {\bibfnamefont {K.}~\bibnamefont
  {Miyoshi}}, \bibinfo {author} {\bibfnamefont {K.}~\bibnamefont {Manami}},
  \bibinfo {author} {\bibfnamefont {R.}~\bibnamefont {Sasai}}, \bibinfo
  {author} {\bibfnamefont {S.}~\bibnamefont {Nishigori}}, \ and\ \bibinfo
  {author} {\bibfnamefont {J.}~\bibnamefont {Takeuchi}},\ }\href {\doibase
  10.1103/PhysRevB.98.195106} {\bibfield  {journal} {\bibinfo  {journal} {Phys.
  Rev. B}\ }\textbf {\bibinfo {volume} {98}},\ \bibinfo {pages} {195106}
  (\bibinfo {year} {2018})}\BibitemShut {NoStop}%
\bibitem [{\citenamefont {Iwaya}\ \emph {et~al.}(2013)\citenamefont {Iwaya},
  \citenamefont {Ogawa}, \citenamefont {Minato}, \citenamefont {Miyoshi},
  \citenamefont {Takeuchi}, \citenamefont {Kuwabara}, \citenamefont {Moriwake},
  \citenamefont {Kim},\ and\ \citenamefont {Hitosugi}}]{Iwaya13}%
  \BibitemOpen
  \bibfield  {author} {\bibinfo {author} {\bibfnamefont {K.}~\bibnamefont
  {Iwaya}}, \bibinfo {author} {\bibfnamefont {T.}~\bibnamefont {Ogawa}},
  \bibinfo {author} {\bibfnamefont {T.}~\bibnamefont {Minato}}, \bibinfo
  {author} {\bibfnamefont {K.}~\bibnamefont {Miyoshi}}, \bibinfo {author}
  {\bibfnamefont {J.}~\bibnamefont {Takeuchi}}, \bibinfo {author}
  {\bibfnamefont {A.}~\bibnamefont {Kuwabara}}, \bibinfo {author}
  {\bibfnamefont {H.}~\bibnamefont {Moriwake}}, \bibinfo {author}
  {\bibfnamefont {Y.}~\bibnamefont {Kim}}, \ and\ \bibinfo {author}
  {\bibfnamefont {T.}~\bibnamefont {Hitosugi}},\ }\href {\doibase
  10.1103/PhysRevLett.111.126104} {\bibfield  {journal} {\bibinfo  {journal}
  {Phys. Rev. Lett.}\ }\textbf {\bibinfo {volume} {111}},\ \bibinfo {pages}
  {126104} (\bibinfo {year} {2013})}\BibitemShut {NoStop}%
\bibitem [{\citenamefont {Pachuta}\ \emph {et~al.}(2019)\citenamefont
  {Pachuta}, \citenamefont {Pentzer},\ and\ \citenamefont
  {Sehirlioglu}}]{Pachuta}%
  \BibitemOpen
  \bibfield  {author} {\bibinfo {author} {\bibfnamefont {K.~G.}\ \bibnamefont
  {Pachuta}}, \bibinfo {author} {\bibfnamefont {E.~B.}\ \bibnamefont
  {Pentzer}}, \ and\ \bibinfo {author} {\bibfnamefont {A.}~\bibnamefont
  {Sehirlioglu}},\ }\href {\doibase 10.1111/jace.16382} {\bibfield  {journal}
  {\bibinfo  {journal} {Journal of the American Ceramic Society}\ }\textbf
  {\bibinfo {volume} {0}} (\bibinfo {year} {2019}),\
  10.1111/jace.16382}\BibitemShut {NoStop}%
\bibitem [{\citenamefont {Masuda}\ \emph {et~al.}(2006)\citenamefont {Masuda},
  \citenamefont {Hamada}, \citenamefont {Seo},\ and\ \citenamefont
  {Koumoto}}]{Masuda06}%
  \BibitemOpen
  \bibfield  {author} {\bibinfo {author} {\bibfnamefont {Y.}~\bibnamefont
  {Masuda}}, \bibinfo {author} {\bibfnamefont {Y.}~\bibnamefont {Hamada}},
  \bibinfo {author} {\bibfnamefont {W.~S.}\ \bibnamefont {Seo}}, \ and\
  \bibinfo {author} {\bibfnamefont {K.}~\bibnamefont {Koumoto}},\ }\href
  {\doibase doi:10.1166/jnn.2006.224} {\bibfield  {journal} {\bibinfo
  {journal} {Journal of Nanoscience and Nanotechnology}\ }\textbf {\bibinfo
  {volume} {6}},\ \bibinfo {pages} {1632} (\bibinfo {year} {2006})}\BibitemShut
  {NoStop}%
\bibitem [{\citenamefont {Pashov}\ \emph {et~al.}(2019)\citenamefont {Pashov},
  \citenamefont {Acharya}, \citenamefont {Lambrecht}, \citenamefont {Jackson},
  \citenamefont {Belashchenko}, \citenamefont {Chantis}, \citenamefont
  {Jamet},\ and\ \citenamefont {van Schilfgaarde}}]{questaal-paper}%
  \BibitemOpen
  \bibfield  {author} {\bibinfo {author} {\bibfnamefont {D.}~\bibnamefont
  {Pashov}}, \bibinfo {author} {\bibfnamefont {S.}~\bibnamefont {Acharya}},
  \bibinfo {author} {\bibfnamefont {W.~R.}\ \bibnamefont {Lambrecht}}, \bibinfo
  {author} {\bibfnamefont {J.}~\bibnamefont {Jackson}}, \bibinfo {author}
  {\bibfnamefont {K.~D.}\ \bibnamefont {Belashchenko}}, \bibinfo {author}
  {\bibfnamefont {A.}~\bibnamefont {Chantis}}, \bibinfo {author} {\bibfnamefont
  {F.}~\bibnamefont {Jamet}}, \ and\ \bibinfo {author} {\bibfnamefont
  {M.}~\bibnamefont {van Schilfgaarde}},\ }\href {\doibase
  https://doi.org/10.1016/j.cpc.2019.107065} {\bibfield  {journal} {\bibinfo
  {journal} {Computer Physics Communications}\ }\textbf {\bibinfo {volume}
  {249}},\ \bibinfo {pages} {107065} (\bibinfo {year} {2019})}\BibitemShut
  {NoStop}%
\bibitem [{com()}]{compdetails}%
  \BibitemOpen
  \href@noop {} {}\bibinfo {note} {Calculations are done in the full-potential
  linearized muffin-tin orbital method.\cite{questaal-paper} Convergence
  parameters were chosen as follows: basis set $spdf-spd$ spherical wave
  envelope functions plus augmented plane waves with a cut-off of 3 Ry,
  augmentation cutoff $l_{max}=4$, {\bf k}-point mesh, $12\times12\times2$. The
  monolayer slabs were separated by a vacuum region of 3 nm.}\BibitemShut
  {Stop}%
\bibitem [{\citenamefont {Atherton}\ \emph {et~al.}(2016)\citenamefont
  {Atherton}, \citenamefont {Butler}, \citenamefont {Taylor}, \citenamefont
  {Hooper}, \citenamefont {Hibbins}, \citenamefont {Sambles},\ and\
  \citenamefont {Mathur}}]{Atherton16}%
  \BibitemOpen
  \bibfield  {author} {\bibinfo {author} {\bibfnamefont {T.~J.}\ \bibnamefont
  {Atherton}}, \bibinfo {author} {\bibfnamefont {C.~A.~M.}\ \bibnamefont
  {Butler}}, \bibinfo {author} {\bibfnamefont {M.~C.}\ \bibnamefont {Taylor}},
  \bibinfo {author} {\bibfnamefont {I.~R.}\ \bibnamefont {Hooper}}, \bibinfo
  {author} {\bibfnamefont {A.~P.}\ \bibnamefont {Hibbins}}, \bibinfo {author}
  {\bibfnamefont {J.~R.}\ \bibnamefont {Sambles}}, \ and\ \bibinfo {author}
  {\bibfnamefont {H.}~\bibnamefont {Mathur}},\ }\href {\doibase
  10.1103/PhysRevB.93.125106} {\bibfield  {journal} {\bibinfo  {journal} {Phys.
  Rev. B}\ }\textbf {\bibinfo {volume} {93}},\ \bibinfo {pages} {125106}
  (\bibinfo {year} {2016})}\BibitemShut {NoStop}%
\bibitem [{SM()}]{SM}%
  \BibitemOpen
  \href@noop {} {}\bibinfo {note} {Supplemental material contains information
  on the DFT results for various Li locations on the surface unit cell, the
  band structures for Na and Be covered CoO$_2$ monolayers, and partial
  occupation with Li on both surfaces. It also contains details on the
  entanglement spectrum results and the numerical charge evaluation in the
  tight-binding model.}\BibitemShut {Stop}%
\bibitem [{\citenamefont {Obana}\ \emph {et~al.}(2019)\citenamefont {Obana},
  \citenamefont {Liu},\ and\ \citenamefont {Wakabayashi}}]{obana}%
  \BibitemOpen
  \bibfield  {author} {\bibinfo {author} {\bibfnamefont {D.}~\bibnamefont
  {Obana}}, \bibinfo {author} {\bibfnamefont {F.}~\bibnamefont {Liu}}, \ and\
  \bibinfo {author} {\bibfnamefont {K.}~\bibnamefont {Wakabayashi}},\ }\href
  {\doibase 10.1103/PhysRevB.100.075437} {\bibfield  {journal} {\bibinfo
  {journal} {Phys. Rev. B}\ }\textbf {\bibinfo {volume} {100}},\ \bibinfo
  {pages} {075437} (\bibinfo {year} {2019})}\BibitemShut {NoStop}%
\bibitem [{\citenamefont {Pershoguba}\ and\ \citenamefont
  {Yakovenko}(2012)}]{Pershoguba12}%
  \BibitemOpen
  \bibfield  {author} {\bibinfo {author} {\bibfnamefont {S.~S.}\ \bibnamefont
  {Pershoguba}}\ and\ \bibinfo {author} {\bibfnamefont {V.~M.}\ \bibnamefont
  {Yakovenko}},\ }\href {\doibase 10.1103/PhysRevB.86.075304} {\bibfield
  {journal} {\bibinfo  {journal} {Phys. Rev. B}\ }\textbf {\bibinfo {volume}
  {86}},\ \bibinfo {pages} {075304} (\bibinfo {year} {2012})}\BibitemShut
  {NoStop}%
\bibitem [{\citenamefont {Mong}\ and\ \citenamefont {Shivamoggi}(2011)}]{Mong}%
  \BibitemOpen
  \bibfield  {author} {\bibinfo {author} {\bibfnamefont {R.~S.~K.}\
  \bibnamefont {Mong}}\ and\ \bibinfo {author} {\bibfnamefont {V.}~\bibnamefont
  {Shivamoggi}},\ }\href {\doibase 10.1103/PhysRevB.83.125109} {\bibfield
  {journal} {\bibinfo  {journal} {Phys. Rev. B}\ }\textbf {\bibinfo {volume}
  {83}},\ \bibinfo {pages} {125109} (\bibinfo {year} {2011})}\BibitemShut
  {NoStop}%
\bibitem [{\citenamefont {Fidkowski}(2010)}]{Fidkowski}%
  \BibitemOpen
  \bibfield  {author} {\bibinfo {author} {\bibfnamefont {L.}~\bibnamefont
  {Fidkowski}},\ }\href {\doibase 10.1103/PhysRevLett.104.130502} {\bibfield
  {journal} {\bibinfo  {journal} {Phys. Rev. Lett.}\ }\textbf {\bibinfo
  {volume} {104}},\ \bibinfo {pages} {130502} (\bibinfo {year}
  {2010})}\BibitemShut {NoStop}%
\bibitem [{\citenamefont {Turner}\ \emph {et~al.}(2010)\citenamefont {Turner},
  \citenamefont {Zhang},\ and\ \citenamefont {Vishwanath}}]{Turner10}%
  \BibitemOpen
  \bibfield  {author} {\bibinfo {author} {\bibfnamefont {A.~M.}\ \bibnamefont
  {Turner}}, \bibinfo {author} {\bibfnamefont {Y.}~\bibnamefont {Zhang}}, \
  and\ \bibinfo {author} {\bibfnamefont {A.}~\bibnamefont {Vishwanath}},\
  }\href {\doibase 10.1103/PhysRevB.82.241102} {\bibfield  {journal} {\bibinfo
  {journal} {Phys. Rev. B}\ }\textbf {\bibinfo {volume} {82}},\ \bibinfo
  {pages} {241102} (\bibinfo {year} {2010})}\BibitemShut {NoStop}%
\bibitem [{\citenamefont {Calabrese}(2016)}]{Calabrese_2016}%
  \BibitemOpen
  \bibfield  {author} {\bibinfo {author} {\bibfnamefont {P.}~\bibnamefont
  {Calabrese}},\ }\href {\doibase 10.1088/1751-8113/49/42/421001} {\bibfield
  {journal} {\bibinfo  {journal} {Journal of Physics A: Mathematical and
  Theoretical}\ }\textbf {\bibinfo {volume} {49}},\ \bibinfo {pages} {421001}
  (\bibinfo {year} {2016})}\BibitemShut {NoStop}%
\bibitem [{\citenamefont {Alexandradinata}\ \emph {et~al.}(2011)\citenamefont
  {Alexandradinata}, \citenamefont {Hughes},\ and\ \citenamefont
  {Bernevig}}]{Alexandradinata}%
  \BibitemOpen
  \bibfield  {author} {\bibinfo {author} {\bibfnamefont {A.}~\bibnamefont
  {Alexandradinata}}, \bibinfo {author} {\bibfnamefont {T.~L.}\ \bibnamefont
  {Hughes}}, \ and\ \bibinfo {author} {\bibfnamefont {B.~A.}\ \bibnamefont
  {Bernevig}},\ }\href {\doibase 10.1103/PhysRevB.84.195103} {\bibfield
  {journal} {\bibinfo  {journal} {Phys. Rev. B}\ }\textbf {\bibinfo {volume}
  {84}},\ \bibinfo {pages} {195103} (\bibinfo {year} {2011})}\BibitemShut
  {NoStop}%
\bibitem [{\citenamefont {Yang}\ \emph {et~al.}(2007)\citenamefont {Yang},
  \citenamefont {Wang},\ and\ \citenamefont {Ding}}]{Yang07}%
  \BibitemOpen
  \bibfield  {author} {\bibinfo {author} {\bibfnamefont {H.-B.}\ \bibnamefont
  {Yang}}, \bibinfo {author} {\bibfnamefont {Z.}~\bibnamefont {Wang}}, \ and\
  \bibinfo {author} {\bibfnamefont {H.}~\bibnamefont {Ding}},\ }\href {\doibase
  10.1088/0953-8984/19/35/355004} {\bibfield  {journal} {\bibinfo  {journal}
  {Journal of Physics: Condensed Matter}\ }\textbf {\bibinfo {volume} {19}},\
  \bibinfo {pages} {355004} (\bibinfo {year} {2007})}\BibitemShut {NoStop}%
\bibitem [{\citenamefont {Yang}\ \emph {et~al.}(2005)\citenamefont {Yang},
  \citenamefont {Pan}, \citenamefont {Sekharan}, \citenamefont {Sato},
  \citenamefont {Souma}, \citenamefont {Takahashi}, \citenamefont {Jin},
  \citenamefont {Sales}, \citenamefont {Mandrus}, \citenamefont {Fedorov},
  \citenamefont {Wang},\ and\ \citenamefont {Ding}}]{Yang05}%
  \BibitemOpen
  \bibfield  {author} {\bibinfo {author} {\bibfnamefont {H.-B.}\ \bibnamefont
  {Yang}}, \bibinfo {author} {\bibfnamefont {Z.-H.}\ \bibnamefont {Pan}},
  \bibinfo {author} {\bibfnamefont {A.~K.~P.}\ \bibnamefont {Sekharan}},
  \bibinfo {author} {\bibfnamefont {T.}~\bibnamefont {Sato}}, \bibinfo {author}
  {\bibfnamefont {S.}~\bibnamefont {Souma}}, \bibinfo {author} {\bibfnamefont
  {T.}~\bibnamefont {Takahashi}}, \bibinfo {author} {\bibfnamefont
  {R.}~\bibnamefont {Jin}}, \bibinfo {author} {\bibfnamefont {B.~C.}\
  \bibnamefont {Sales}}, \bibinfo {author} {\bibfnamefont {D.}~\bibnamefont
  {Mandrus}}, \bibinfo {author} {\bibfnamefont {A.~V.}\ \bibnamefont
  {Fedorov}}, \bibinfo {author} {\bibfnamefont {Z.}~\bibnamefont {Wang}}, \
  and\ \bibinfo {author} {\bibfnamefont {H.}~\bibnamefont {Ding}},\ }\href
  {\doibase 10.1103/PhysRevLett.95.146401} {\bibfield  {journal} {\bibinfo
  {journal} {Phys. Rev. Lett.}\ }\textbf {\bibinfo {volume} {95}},\ \bibinfo
  {pages} {146401} (\bibinfo {year} {2005})}\BibitemShut {NoStop}%
\bibitem [{\citenamefont {Qian}\ \emph {et~al.}(2006)\citenamefont {Qian},
  \citenamefont {Wray}, \citenamefont {Hsieh}, \citenamefont {Viciu},
  \citenamefont {Cava}, \citenamefont {Luo}, \citenamefont {Wu}, \citenamefont
  {Wang},\ and\ \citenamefont {Hasan}}]{Qian06}%
  \BibitemOpen
  \bibfield  {author} {\bibinfo {author} {\bibfnamefont {D.}~\bibnamefont
  {Qian}}, \bibinfo {author} {\bibfnamefont {L.}~\bibnamefont {Wray}}, \bibinfo
  {author} {\bibfnamefont {D.}~\bibnamefont {Hsieh}}, \bibinfo {author}
  {\bibfnamefont {L.}~\bibnamefont {Viciu}}, \bibinfo {author} {\bibfnamefont
  {R.~J.}\ \bibnamefont {Cava}}, \bibinfo {author} {\bibfnamefont {J.~L.}\
  \bibnamefont {Luo}}, \bibinfo {author} {\bibfnamefont {D.}~\bibnamefont
  {Wu}}, \bibinfo {author} {\bibfnamefont {N.~L.}\ \bibnamefont {Wang}}, \ and\
  \bibinfo {author} {\bibfnamefont {M.~Z.}\ \bibnamefont {Hasan}},\ }\href
  {\doibase 10.1103/PhysRevLett.97.186405} {\bibfield  {journal} {\bibinfo
  {journal} {Phys. Rev. Lett.}\ }\textbf {\bibinfo {volume} {97}},\ \bibinfo
  {pages} {186405} (\bibinfo {year} {2006})}\BibitemShut {NoStop}%
\bibitem [{\citenamefont {Shimojima}\ \emph {et~al.}(2006)\citenamefont
  {Shimojima}, \citenamefont {Ishizaka}, \citenamefont {Tsuda}, \citenamefont
  {Kiss}, \citenamefont {Yokoya}, \citenamefont {Chainani}, \citenamefont
  {Shin}, \citenamefont {Badica}, \citenamefont {Yamada},\ and\ \citenamefont
  {Togano}}]{Shimojima06}%
  \BibitemOpen
  \bibfield  {author} {\bibinfo {author} {\bibfnamefont {T.}~\bibnamefont
  {Shimojima}}, \bibinfo {author} {\bibfnamefont {K.}~\bibnamefont {Ishizaka}},
  \bibinfo {author} {\bibfnamefont {S.}~\bibnamefont {Tsuda}}, \bibinfo
  {author} {\bibfnamefont {T.}~\bibnamefont {Kiss}}, \bibinfo {author}
  {\bibfnamefont {T.}~\bibnamefont {Yokoya}}, \bibinfo {author} {\bibfnamefont
  {A.}~\bibnamefont {Chainani}}, \bibinfo {author} {\bibfnamefont
  {S.}~\bibnamefont {Shin}}, \bibinfo {author} {\bibfnamefont {P.}~\bibnamefont
  {Badica}}, \bibinfo {author} {\bibfnamefont {K.}~\bibnamefont {Yamada}}, \
  and\ \bibinfo {author} {\bibfnamefont {K.}~\bibnamefont {Togano}},\ }\href
  {\doibase 10.1103/PhysRevLett.97.267003} {\bibfield  {journal} {\bibinfo
  {journal} {Phys. Rev. Lett.}\ }\textbf {\bibinfo {volume} {97}},\ \bibinfo
  {pages} {267003} (\bibinfo {year} {2006})}\BibitemShut {NoStop}%
\bibitem [{\citenamefont {Hong}\ \emph {et~al.}(2019)\citenamefont {Hong},
  \citenamefont {Hu}, \citenamefont {Freeland}, \citenamefont {Cabana},
  \citenamefont {{\"O}g{\"u}t},\ and\ \citenamefont {Klie}}]{Hong19}%
  \BibitemOpen
  \bibfield  {author} {\bibinfo {author} {\bibfnamefont {L.}~\bibnamefont
  {Hong}}, \bibinfo {author} {\bibfnamefont {L.}~\bibnamefont {Hu}}, \bibinfo
  {author} {\bibfnamefont {J.~W.}\ \bibnamefont {Freeland}}, \bibinfo {author}
  {\bibfnamefont {J.}~\bibnamefont {Cabana}}, \bibinfo {author} {\bibfnamefont
  {S.}~\bibnamefont {{\"O}g{\"u}t}}, \ and\ \bibinfo {author} {\bibfnamefont
  {R.~F.}\ \bibnamefont {Klie}},\ }\href {\doibase 10.1021/acs.jpcc.8b11661}
  {\bibfield  {journal} {\bibinfo  {journal} {The Journal of Physical Chemistry
  C}\ }\textbf {\bibinfo {volume} {123}},\ \bibinfo {pages} {8851} (\bibinfo
  {year} {2019})}\BibitemShut {NoStop}%
\end{thebibliography}%

\end{document}